\documentclass[aps,prb,twocolumn,amsmath,amssymb,showpacs]{revtex4}
\usepackage{graphicx}
\usepackage{dcolumn}
\usepackage{bm}
\begin{document}

\title{Landau levels in a topological insulator}

\author{P. Schwab, M. Dzierzawa}
\affiliation{Institut f\"ur Physik, Universit\"at Augsburg, 86135 Augsburg, Germany
}

\date{\today}

\begin{abstract}
Two recent experiments successfully observed Landau levels in the tunneling spectra of the topological insulator
Bi$_2$Se$_3$.
To mimic the influence of a scanning tunneling microscope tip on the Landau levels 
we solve the two-dimensional Dirac equation in the presence of a 
localized electrostatic potential. We find that the STM tip not only shifts the Landau levels, 
but also suppresses for a realistic choice of parameters the negative branch of Landau levels.
\end{abstract}

\maketitle

Topological insulators have recently attracted considerable interest,
especially after their experimental discovery in two- and three dimensions
\cite{moore2009,hasan2010,konig2007,hsieh2008,hsieh2009a,hsieh2009,chen2009,xia2009,kuroda2010}.
While insulating in the bulk, such materials possess gapless 
edge states whose existence depends on time reversal invariance
\cite{kane2005, kane2005a,bernevig2006,moore2007,fu2007,murakami2007,roy2009}.
This makes the latter robust against time-reversal symmetric perturbations
such as impurity scattering and at the same time very sensitive to 
time-reversal breaking ones such as magnetic fields.
When the topological insulator is a three-dimensional system, the gapless excitations are confined 
to its surface and form a two-dimensional conductor.

The Landau quantization of the surface states of the three-dimensional topological insulator Bi$_2$Se$_3$
has been reported in two recent experiments \cite{cheng2010,hanaguri2010}. In both cases the Landau levels were
detected using a scanning tunneling microscope (STM). 
The differential tunneling conductance $\partial I / \partial V$ measures the local density of states of electrons at energy $eV$.
In the absence of magnetic fields the measured spectra are
consistent with a Dirac dispersion of the surface states. 
In the presence of a magnetic field Landau levels are expected at energies
\begin{equation}
E_n = E_{D} +  {\rm sgn} (n) v_F \sqrt{2 e \hbar |n| B  }
\end{equation}
where $E_{D}$ is the energy of the Dirac point, $n=0, \pm 1, \pm 2, \dots$ is the Landau level index, $v_F$ is the velocity, 
$e$ is the unit charge, $\hbar$ is the Planck constant, and $B$ is the magnetic field.
A series of unequally spaced Landau levels has been observed in the above-mentioned experiments, including a $B$-independent level at 
the Dirac point. However, while the theory predicts a positive ($n > 0$) and a negative ($n < 0 $) branch of Landau levels 
experimentally only the positive branch has been seen. 
It has been speculated  \cite{cheng2010,hanaguri2010} that the absence of Landau levels
below the Dirac point may result form an overlapping of the surface states with the bulk valence band. On the other hand, as pointed out
in Ref. [\onlinecite{hanaguri2010}], 
near the Fermi energy the bulk conduction band overlaps with the surface state, but still Landau levels are clearly
observed.

In this paper we investigate another possible reason for the absence of the negative branch of Landau levels, 
namely the electrostatic effect due to the STM tip:  
The authors of Ref.~[\onlinecite{cheng2010}] have noticed that the Dirac point in their STM measurements is about 200 meV below the Fermi level,
while the Fermi level determined by angular-resolved photoemission spectroscopy is only 120 meV above the Dirac point \cite{zhang2010}, 
and suggested that this discrepancy might be due to the electrostatic interaction between the STM tip and the sample.
We will elaborate this idea further and will demonstrate that such a potential may indeed strongly suppress the negative branch of the Landau levels.

In the following we will present the model under investigation, sketch the numerical methods and finally we will present the results.
Close to the Dirac point the surface states of a topological insulator with a single Dirac cone can be described by the
Hamiltonian
\cite{zhang2009,fu2009,hasan2010}
\begin{equation} \label{eq2}
H=  v_F ({ \bf p} + e {\bf A} )   \cdot {\boldsymbol \sigma}  + V({\bf x} ),
\end{equation}
where ${\bf p}$ is the two-dimensional momentum operator, ${\bf A}$ is the vector potential and ${\boldsymbol\sigma}$ 
are the Pauli matrices.
Note that ${\boldsymbol \sigma}$ is in general not the spin operator. For example, for Bi$_2$Se$_3$, symmetry requires
for the spin operator the relation
${\bf S} \propto {\bf e}_z \times  {\boldsymbol \sigma}$. \cite{fu2009} 
$V({\bf x})$ is the electrostatic potential caused by the tip, which 
we characterize by its depth and width. 
The depth can be extracted from the experiment from the position of the Dirac point.
The width is not directly known but can be estimated to be of the order 10 -- 20 nm. \cite{xi2011}
The results presented later are obtained with a Gaussian potential,
$V({\bf x})  = V_0 \exp( - |{\bf x} |^2 / 2 \sigma^2 )$. 

\begin{figure}
\includegraphics[width=0.45\textwidth]{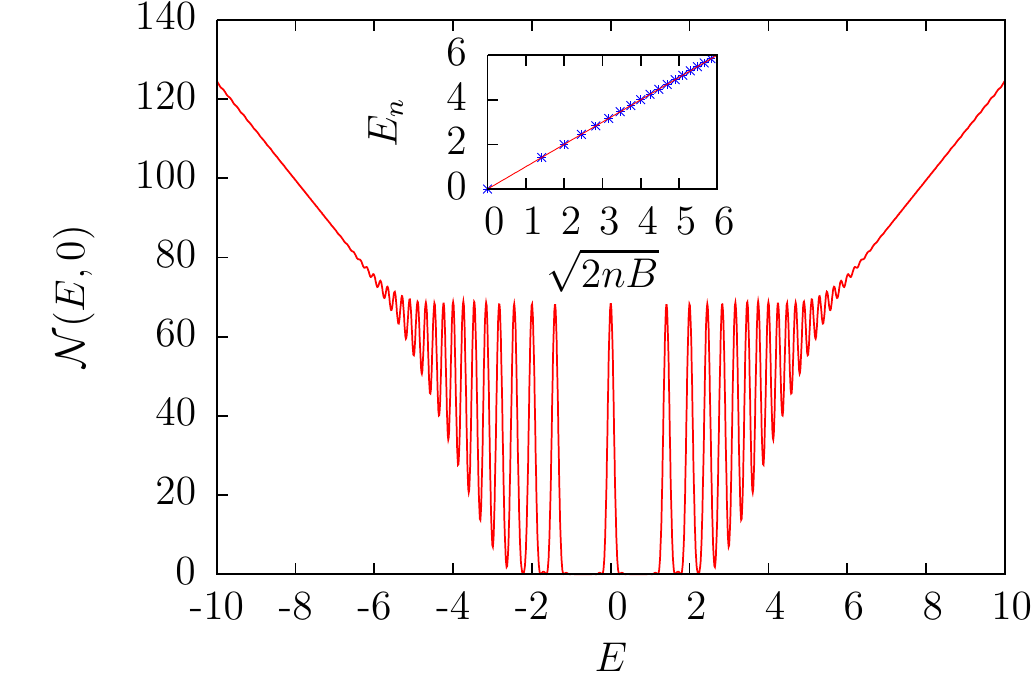}
\caption{  \label{fig1}
Local density of states (arbitrary units) as a function of energy
for the Hamiltonian (\ref{eq2}) in the absence of a potential $V({\bf x} )$ but in the presence of a
magnetic field $B= B_0$. 
The energy is measured in units of
$E_0 = \hbar v_F /x_0$.
The inset shows the energy $E$ as a function of the Landau level index $n$ (counting starts at $E=0$).
}
\end{figure}

\begin{figure}
\includegraphics[width=0.45\textwidth]{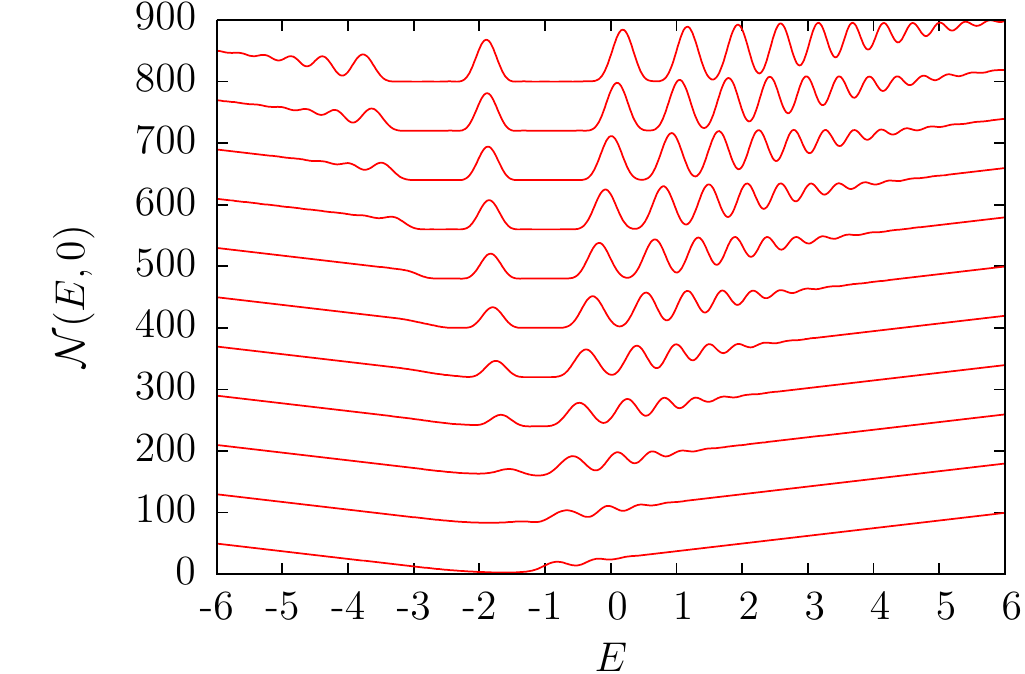}
\caption{  \label{fig2}
Local density of states in the presence of a Gaussian potential with $V_0= -2 E_0$ and $\sigma = 3 x_0$. 
The curves correspond to magnetic field strengths $B/B_0 =0, 0.2 \dots 2 $ (from bottom to top).
With the parameters given in the text
the energy scale is $E_0 \approx 33$ meV and the magnetic field scale is
$B_0 \approx 6.6$ T. The potential has a depth of 66 meV and a width of 30 nm. 
}
\end{figure}

\begin{figure}
\includegraphics[width=0.45\textwidth]{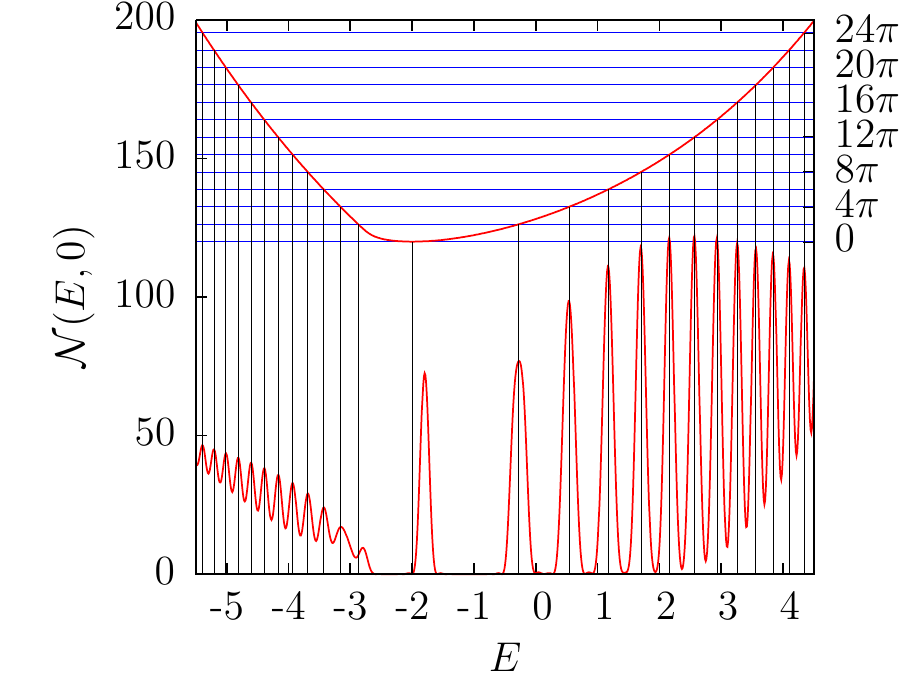}
\caption{  \label{fig3}
Semiclassical analysis of the Landau levels for $B=B_0$ in the presence of a potential $V({\bf x }) $.  The upper curve shows 
the action $I/\hbar$ for closed trajectories starting and ending ${\bf x} = 0 $ (right axis), the lower curve shows the local density of states
${\cal N} (E, {\bf x}\! = \! 0 )$ (left axis). The depth and width of the potential $V({\bf x})$ are the same as in Fig. \ref{fig2}.
}
\end{figure}

\begin{figure}[t]
\includegraphics[width=0.45\textwidth]{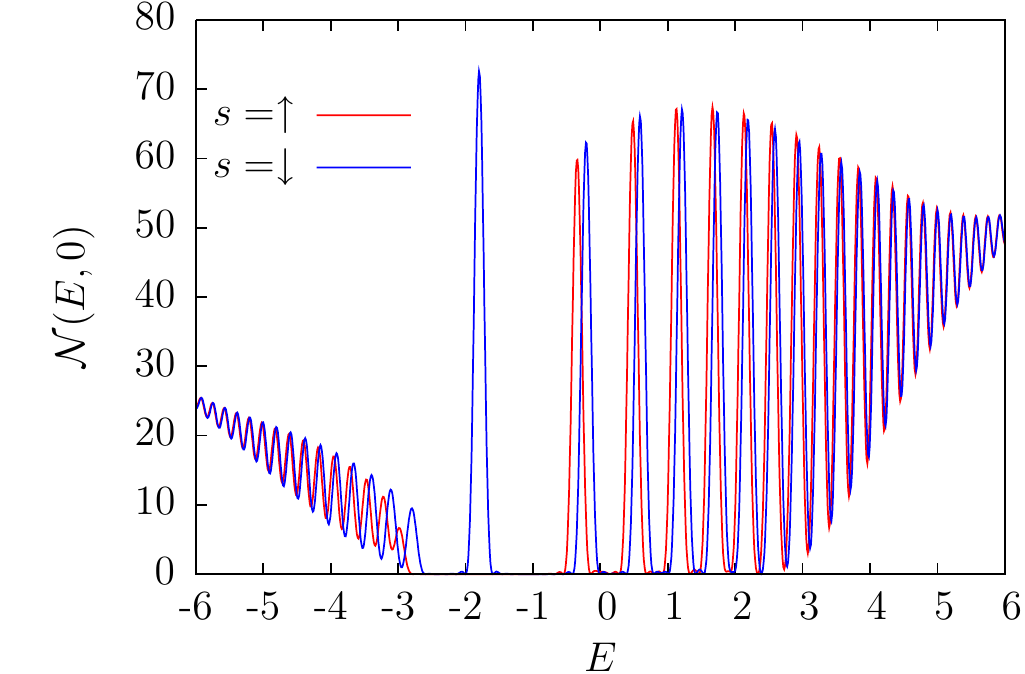}
\caption{  \label{fig4}
Spin resolved density of states ${\cal N}_s(E, {\bf x}\!  = \!  0   )$ for the same parameters as in Fig. \ref{fig3}. 
}
\end{figure}

For the numerical treatment it is convenient to introduce dimensionless units. In the following we 
measure distances in units of an arbitrary length scale $x_0$.  
The energy is measured in units of $E_0 = \hbar v_F /x_0$, and the magnetic
field in units of $B_0 = \hbar/(e x_0^2)$. 
Numerically a system of size $L \times L$ is investigatied. The results presented in Figs. \ref{fig1}--\ref{fig4} are obtained with $L=10 \pi x_0$
using periodic boundary conditions.
We expand the wave function as
\begin{equation}
\Psi({ \bf x }) = \sum_{\bf k } {\rm e}^{ i {\bf k } \cdot {\bf  x } } \Psi({\bf k} ), \quad {\bf k } =  \frac{2 \pi}{L} (n_x, n_y ) 
\end{equation} 
and truncate the expansion, $ |n_x|$, $|n_y| \leq N $; empirically we found that $N \approx 100$ 
is large enough for our purposes.
For the vector potential we use the gauge ${\bf A} = (0, Bx, 0 )$ with a discontinuity at $x = \pm L/2$.
The quantity to be calculated is the 
(spin dependent) local density of states defined as
\begin{eqnarray}
{\cal N}_s(\epsilon,  { \bf x} )  &=  &\sum_\lambda \Psi^*_\lambda(s, {\bf x} ) \Psi_\lambda(s, {\bf x} ) \delta (\epsilon - \epsilon_\lambda ), \\
{\cal N}(\epsilon, {\bf x}  )& = &{\cal N}_\uparrow(\epsilon,  { \bf x} )+  {\cal N}_\downarrow(\epsilon,  { \bf x} )
.\end{eqnarray}
The spinors $\Psi_\lambda(s, {\bf x } ) $ are eigenfunctions and $\epsilon_\lambda$ are eigenvalues of the Hamiltonian (\ref{eq2}).
To evaluate the density of states  at ${\bf x} = 0 $ we apply a method based on an expansion in Chebyshev polynomials following 
Ref. [\onlinecite{weisse2006}].
Instead of a sequence of $\delta$-functions the Chebyshev expansion produces a broadened density of states, 
the broadening being determined by the number of polynomials kept in the expansion. 

The accuracy of the method is demonstrated in Fig. \ref{fig1}, where we show the local density of states 
in the presence of a magnetic field ($B= B_0$) but in the absence of the potential $V({\bf x } )$.
At low energies one observes a sequence of Landau levels at positions
$E_n = \pm E_0 \sqrt{ 2 n B/B_0}$ as expected from the analytical calculations. The spectrum is reproduced with a very high accuracy
as it is shown in the inset of Fig.~\ref{fig1}.
Due to the truncation of the Chebyshev expansion the Landau levels in our figure have a finite width. 
As a consequence only a few discrete peaks are seen in the low energy region of the density of states, 
whereas  at larger energy the Landau levels overlap such that one observes the linear density of states of the 
Dirac Hamiltonian. 

Figure \ref{fig2} depicts the local density of states in the presence of a Gaussian potential with
 $V_0 = - 2 E_0$ and $\sigma = 3 x_0$.
The magnetic field varies from $B=0$ to $B=2B_0$ in steps of $\Delta B = 0.2 B_0$. 
For clarity we use offsets for the different magnetic field strengths.
With $x_0 = 10$ nm and $v_F = 5 \times 10^5$ m/s the energy and magnetic field scales are  
$E_0 \approx 33$ meV and $B_0 \approx 6.6$ T. With our choice of parameters we are close to the values given in 
Refs. [\onlinecite{cheng2010}] and [\onlinecite{hanaguri2010}].
For zero magnetic field (lowest curve), 
the potential shifts the minimum of the density of states to a lower energy, and the density of states is no longer symmetric
around the minimum.
In the presence of a magnetic field
one observes several Landau levels.
However, the negative branch of the Landau levels is suppressed. Only for large magnetic field
peaks that correspond to negative Landau level index $n$ reappear.

In order to obtain a qualitative understanding of these findings 
we present now a semiclassical analysis of the effect.
We start with the classical equations of motion 
for wave packets formed by the eigenstates of the Hamiltonian (\ref{eq2}),
\begin{eqnarray}
{\bf v}  & =  & \pm v_F {\hat {\bf k } }  \\
\hbar \dot {\bf k} & =  & - e \, {\bf v} \times {\bf B} - \nabla V({\bf x } ) 
;\end{eqnarray}
here ${\bf v}$ is the velocity, ${\hat {\bf k }} $ is the direction of ${\bf k }$ and the plus and minus sign correspond
to particles and holes respectively.
In the absence of the potential $V({\bf x} )$ 
the magnetic force constrains the trajectories to circles with cyclotron radius $r_c = \hbar k / e B $.
Using the Bohr-Sommerfeld quantization condition
\begin{equation}
I= \oint {\bf p } \cdot d {\bf x}  = 2 \pi \hbar ( n + \gamma ), \quad {\bf p } = \hbar {\bf k } - e {\bf A}
\end{equation}
with $\gamma = 0$
the correct Landau level spectrum is recovered. 
Generally, $\gamma$ is the sum of a Berry phase and a Maslov contribution which cancel each other in the case of massless Dirac fermions,
see for example Ref. [\onlinecite{fuchs2010}] for a recent discussion.

In the presence of the electrostatic potential $V({\bf x})$ there is a competition between magnetic
and electric forces.
Considering for example a circular motion around the origin we find a modified cyclotron radius,
\begin{equation}
r_c = \frac{ \hbar v_F k }{ e v_F B \pm V'(r_c)}
.\end{equation}
For a Gaussian potential with negative $V_0$ the derivative $V'(r_c)$ is positive 
and thus the potential reduces the cyclotron radius for electrons (due to the extra force directing towards the center of the 
cyclotron orbit) but enlarges the cyclotron radius for holes.
An analogous behavior is found for the action integral $I$: for a given $k$,  the potential increases the action of closed loops for holes
and decreases the action for the electrons.

The upper part of Fig. \ref{fig3} depicts numerical results for the action $I$ (divided by $\hbar$) 
for trajectories that start and end at ${\bf x} = 0 $ 
as a function of energy. 
The trajectories are closed, in general they are not periodic but form rosette-like orbits.
Clearly for electrons $(E > V_0= -2 E_0)$ the action grows much slower as a function
of energy than for the holes just as we argued before for circular trajectories.
In the lower part of Fig. \ref{fig3} we show again the local density of states at ${\bf x} = 0 $, but compared to Fig. \ref{fig2} we increased
the number of Chebyshev polynomials such that the peaks become sharper. Due to this improved resolution the density of states below 
$E \approx - 2 E_0$ is no longer smooth but also shows a pronounced peak structure. Furthermore, with exception of the two peaks at 
$E \approx - 2 E_0$ and $E\approx -3 E_0$, the Bohr-Sommerfeld quantization condition accurately reproduces the peak positions.

Figure \ref{fig4} shows the spin resolved density of states for the  parameters of Fig. \ref{fig3}.
The central peak is fully spin-polarized which is not a surprise since already in the absence of $V({\bf x} )$ the $n=0$ Landau
level is spin-polarized.
More surprisingly
the electrostatic potential $V({\bf x} ) $
splits the higher Landau levels into two spin polarized peaks, an effect for which we do not have a semiclassical 
explanation at the moment. 

%

In summary we investigated the Landau levels in the surface states of a topological insulator. In order to mimic the influence of
an STM tip on the local density of states we
included an electrostatic potential in the description. For a realistic set of parameters the local density of states behaves
very similar to what has been observed experimentally, namely the negative branch of Landau levels appears to be suppressed.
The origin of the effect is the widening of the cyclotron orbits of the holes due to the electric force. 
We notice that similar  STM studies exist also for the Landau levels in graphene,
 where the low energy physics is governed by Dirac cones as well. 
In a study of graphene on graphite, where metallic graphite can screen the electrostatic effects due to the STM tip,
both branches of Landau levels have been observed
\cite{li2009}. On the other hand in graphene on insulating SiO$_2$ substrates there is no such screening and only one branch 
of Landau levels is observed experimentally\cite{luican2011}, similar to what was found for Bi$_2$Se$_3$.
This suggests that in both cases the electrostatic field due to the STM tip is the origin of the asymmetric STM spectra.

We thank the Deutsche Forschungsgemeinschaft (SPP1285) for financial support.
\bibliography{bib}
\end{document}